\documentclass[10pt, preprintnumbers, aps, prd, twocolumn, superscriptaddress, altaffilletter, nofootinbib, bibnotes]{revtex4-2}

\usepackage{amssymb,amsmath,mathrsfs,enumerate}
\usepackage{bm}
\usepackage{graphicx}
\usepackage{feynmp-auto}
\usepackage{mathtools}
\usepackage{caption}
\usepackage{subcaption}
\usepackage[dvipsnames]{xcolor}
\usepackage[colorlinks=true, linkcolor=blue, urlcolor=Blue, citecolor=Mulberry]{hyperref}
\usepackage{orcidlink}
\usepackage{multirow}
\usepackage{booktabs}
\usepackage[normalem]{ulem}
\usepackage{microtype}  
\usepackage{ragged2e}   

\makeatletter
\patchcmd{\@makecaption}{\ignorespaces}{\justifying\ignorespaces}{}{}
\makeatother

\begin{document}

\font\mini=cmr10 at 0.8pt
\title{Not so good $\nu$s for Higgsino dark matter as LZ excess:\\
stringent limits from Super-Kamiokande and IceCube}

\author{Debajit Bose \orcidlink{0000-0001-8594-8885}}
\email{debajitbose550@gmail.com}
\affiliation{Centre for High Energy Physics, Indian Institute of Science, C. V. Raman Avenue, Bengaluru 560012, India}
\author{Akash Kumar Saha \orcidlink{0000-0002-3033-2589}}
\email{akashks@iisc.ac.in}
\affiliation{Centre for High Energy Physics, Indian Institute of Science, C. V. Raman Avenue, Bengaluru 560012, India}
\author{Deep Jyoti Das \orcidlink{0009-0008-1293-0105}}
\email{deepjyoti@iisc.ac.in}
\affiliation{Centre for High Energy Physics, Indian Institute of Science, C. V. Raman Avenue, Bengaluru 560012, India}
\author{Rinchen Sherpa~\orcidlink{0009-0008-4611-1044}}
\email{rinchens@iisc.ac.in}
\affiliation{Centre for High Energy Physics, Indian Institute of Science, C. V. Raman Avenue, Bengaluru 560012, India}
\author{Abhijeet Singh \orcidlink{0009-0004-8267-7919}}
\email{abhijeets@iisc.ac.in}
\affiliation{Centre for High Energy Physics, Indian Institute of Science, C. V. Raman Avenue, Bengaluru 560012, India}
\author{Durba Ghosh}
\email{durbaghosh@iisc.ac.in}
\affiliation{Centre for High Energy Physics, Indian Institute of Science, C. V. Raman Avenue, Bengaluru 560012, India}
\author{Jaya Doliya \orcidlink{0009-0003-8407-7442}}
\email{jayadoliya@iisc.ac.in}
\affiliation{Centre for High Energy Physics, Indian Institute of Science, C. V. Raman Avenue, Bengaluru 560012, India}
\author{Subhadip Bouri \orcidlink{0000-0002-4971-8916}}
\email{subhadipb@iisc.ac.in}
\affiliation{Department of Physics, Indian Institute of Science, C. V. Raman Avenue, Bengaluru 560012, India}
\affiliation{Centre for High Energy Physics, Indian Institute of Science, C. V. Raman Avenue, Bengaluru 560012, India}
\author{Biprajit Mondal \orcidlink{0009-0002-0844-2578}}
\email{biprajitm@iisc.ac.in}
\affiliation{Centre for High Energy Physics, Indian Institute of Science, C. V. Raman Avenue, Bengaluru 560012, India}
\author{Ranjini Mondol \orcidlink{0000-0002-1331-0118}}
\email{ranjini12m@gmail.com,\,\,ranjini.mondol@saha.ac.in}
\affiliation{Theory Division, Saha Institute of Nuclear Physics, 1/AF, Bidhannagar, Kolkata 700064, India}
\author{Nirmal Raj \orcidlink{0000-0002-4378-1201}}
\email{nraj@iisc.ac.in}
\affiliation{Centre for High Energy Physics, Indian Institute of Science, C. V. Raman Avenue, Bengaluru 560012, India}
\author{Ranjan Laha \orcidlink{0000-0001-7104-5730}}
\email{ranjanlaha@iisc.ac.in}
\affiliation{Centre for High Energy Physics, Indian Institute of Science, C. V. Raman Avenue, Bengaluru 560012, India}
\date{\today}
%
%
\begin{abstract}

The LUX-ZEPLIN (LZ) collaboration has recently reported a single nuclear recoil event at a high recoil energy of about 250~keV. This has been interpreted as inelastic scattering of dark matter that is a supersymmetric Higgsino with a mass splitting between the neutral states of a few 100 keV. Such dark matter may be captured at high recoil in the Sun through scattering on heavy elements in it, and annihilate to $W^+W^-$ and $ZZ$, in turn giving rise to a neutrino flux detectable on Earth. Using measurements of atmospheric electron- and muon-neutrino fluxes by Super-Kamiokande and IceCube, we constrain thermal and non-thermal Higgsino dark matter, excluding inter-state mass splittings $\lesssim 557$~keV. 
This disfavors Higgsino-like interpretations of the LZ event for standard halo velocities.
\begin{center}
\vspace{-.34cm}{\mini ==================================================================== One of the authors is convinced that this is a background event as it is too close to the upper energy cut ==================================================================.}
\end{center}
\end{abstract}
\maketitle

\section{Introduction}
\label{sec:intro}

The LUX-ZEPLIN (LZ) experiment, seeking direct detection of dark matter (DM), has seen a 248 $\pm$ 23 (stat) $\pm$ 23 (sys) keV nuclear recoil energy event, LZ230616, in its 2.84~tonne-year dataset at a global (local) significance of 2.6$\sigma$ (3.4$\sigma$)~\cite{LZ:2026axp}. The collaboration interpreted this as induced by effective operators that lead to momentum-dependent elastic scattering (via a magnetic moment) or endothermic inelastic scattering; other operators such as scalar-scalar interactions would have produced events at low recoils that were not observed. One particularly attractive explanation of LZ230616~\cite{Fan:2026kxx, Freese:2026sga, Wu:2026nhi, Yin:2026jnn, DiMauro:2026ldr, Du:2026guj} is, with an inelastic mass splitting $\delta \simeq$ 350 keV, the thermal Higgsino, a 1.08~TeV mass supersymmetric partner of the Higgs boson -- arising in weak-scale supersymmetric extensions of the Standard Model (SM)~\cite{Martin:1997ns,Giudice:2004tc,Fox:2014moa} -- that freezes out to the right DM abundance via co-annihilations with slightly heavier neutral and charged Higgsino states, dubbed the ``last-standing electroweak WIMP" and a poster child for electroweak multiplets in the so-called minimal DM paradigm~\cite{Cirelli:2005uq}. Such Higgsino-like dark matter is accessible at collider energies,  is being sought in indirect searches for its annihilation products~\cite{Dessert:2022evk, Rodd:2024qsi, Safdi:2025sfs, Aghaie:2025iyn}, and possibly large $\delta$s may be cornered in compact stars~\cite{McCullough:2010ai, Krall:2017xij, Baryakhtar:2017dbj, Acevedo:2019agu, Alvarez:2023fjj, Acevedo:2024ttq}.

\begin{figure}[t]
    \centering
    \includegraphics[width=0.925\linewidth]{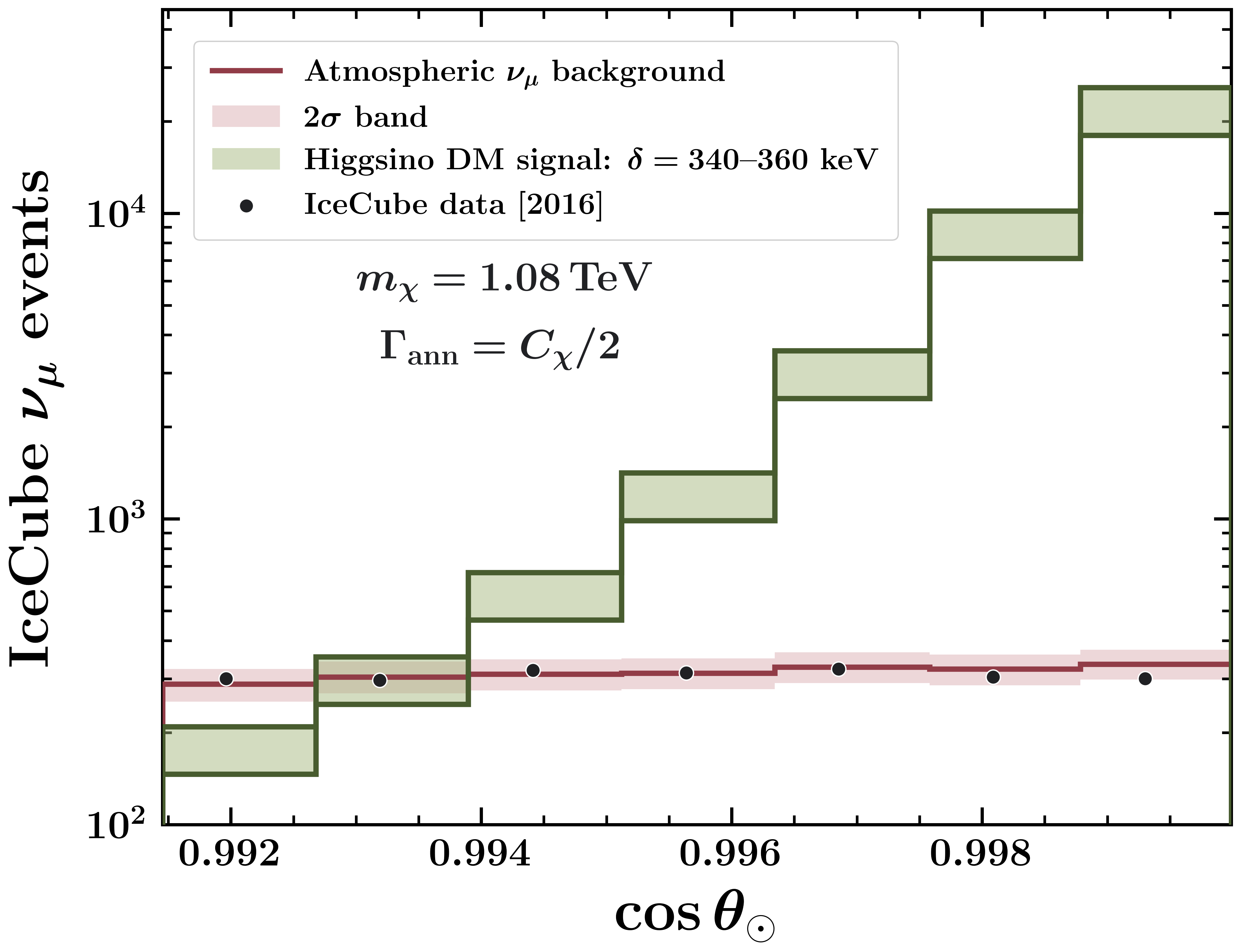}
    \caption{\justifying Solar opening angle event distribution of our $\nu_\mu$ signal at IceCube~\cite{IceCube:2016dgk} from the thermal Higgsino. 
    Our event prediction for inter-state mass splitting $\delta =340-360$~keV, as preferred by the LZ230616 event, clearly exceeds the background and observed event count across several bins and is hence disfavored.
    See Sec.~\ref{sec:datasets} for more details.}
    \label{fig:higgsino_IceCube_numu_equilibrium_Fig2.png}
\end{figure}

While the thermal Higgsino DM interpretation is in slight tension with a lack of events in LZ's high-energy sideband~\cite{Rodd:2026tyn,Dent:2026bji} (which could be tested in LZ's upcoming runs and relaxed by considering a lighter non-thermal Higgsino DM), a more urgent tension has been pointed out. The 1.08 TeV Higgsino DM can be captured in the Sun for $\delta$ $<$ 566~keV by scattering on its constituent heavy elements, e.g., lead, thermalize within the solar core, and subsequently annihilate to $W^+W^-$/ $ZZ$ giving rise to a high-energy neutrino flux that is constrained by IceCube~\cite{Pospelov:2026ewn}. 
This does not as such rule out the thermal Higgsino interpretation of LZ230616, as the event may still have been produced from the high-speed tail of the local DM velocity distribution after accounting for possible boosts due to motion relative to the Large Magellanic Cloud (LMC)~\cite{Fan:2026kxx}. While Ref.~\cite{Pospelov:2026ewn} placed constraints using a 2025 data release of IceCube~\cite{IceCube:2025fcu}, we study the solar capture limits from different measurements: that of atmospheric neutrinos at Super-Kamiokande (Super-K) and other IceCube datasets\,\cite{IceCube:2016dgk, IceCube:2015mgt, Super-Kamiokande:2015qek}, applying angular cuts wherever applicable to pick out signal fluxes from the direction of the Sun. Our strongest limit is $\delta \lesssim 557$~keV, which is comparable to that obtained in Ref.~\cite{Pospelov:2026ewn}. 
We also consider a non-thermal Higgsino to be the DM, for which we estimate limits in the space of $\delta$ and its mass $m_\chi$.
More generally, solar neutrino measurements are being actively studied to constrain DM~\cite{IceCube:2016dgk, IceCube:2021xzo, Maity:2023rez, Bose:2023yll, Krishna:2025ncv, Nguyen:2025ygc, IceCube:2025fcu, Nguyen:2026nhe, Nguyen:2026apa, Bose:2026yun, Nguyen:2026pdr}. 
As Super-K and IceCube cannot discriminate between neutrinos and antineutrinos at these energies, we denote the sum of their contributions as $\nu_\alpha$ for flavor $\alpha$.  
%

\section{Solar capture of inelastic dark matter and the neutrino flux}

The solar capture rate of Higgsino DM (``$\chi$'' hereafter) is (see, e.g., Ref.~\cite{Garani:2017jcj})
\begin{eqnarray}
    C_\chi
    &=& 
 \nonumber   \sum_A
    \int_0^{R_\odot} 4\pi r^2\,dr\, n_A(r)
    \int_0^{u_{\chi, \max}} du_\chi\,
    \frac{d\Phi_\chi}{du_\chi}
    \\
\nonumber  &&   \times
    \int_{E_R^{\min}}^{E_R^+}
    dE_R\,
    \frac{d\sigma_{\chi A}}{dE_R}~, \\
\nonumber  \frac{d {\Phi_\chi}}{du_\chi}&=&\frac{\rho_{\chi}}{m_{\chi}}F(u_\chi)\frac{w^2(r,u_\chi)}{u_\chi}~,    \\
\nonumber   F(u_\chi) &\equiv& u_\chi^2\int d \Omega_{u_\chi} f_{\rm MB}(\mathbf{u}_{\bm{\chi}}+\mathbf{v}_\odot)~, \\
 f_{\rm MB}(u_\chi)&=& \frac{1}{N_{\rm esc}(\pi u_0^2)^{3/2}} e^{-u_\chi^2/u_0^2} \Theta(u_{\rm esc}^{\rm gal} - u_\chi) \, ,
\label{eq:capture_rate}
\end{eqnarray}
where the local DM density $\rho_{\chi}=0.4 \, {\rm GeV/cm^3}$, $u_{\chi, \max}$ is the maximum asymptotic DM speed that allows capture at radius $r$, $E^{\rm{min}}_R=\mathrm{max}[E_R^-, \frac{1}{2}m_{\chi}u^2-\delta]$ is the minimum recoil energy required for capture,  $\mathbf{u_\chi}$ is the halo DM velocity and $f_{\rm MB}$ a truncated Maxwell-Boltzmann distribution\footnote{For deviations from this distribution and their consequences, see Ref.~\cite{Bose:2022ola}.}, with the circular speed $u_0=220~{\rm km/s}$, Galactic escape speed $u_{\rm esc}^{\rm gal}=544~{\rm km/s}$, Sun's speed in the Galactic halo $v_\odot\simeq232~{\rm km/s}$, and $N_{\rm esc}$ is the normalization constant; $w(r,u_\chi)=\sqrt{u_\chi^2+v_{\rm esc}^2(r)}$ and $F(u_\chi)$ is the angle-averaged speed distribution of DM at the Sun.

$\chi$ is bound to the Sun's gravitational potential if its scattering on solar targets depletes sufficient kinetic energy. For inelastic scattering on nuclei, $\chi + A \rightarrow \tilde{\chi} + A$ ($\tilde{\chi}$ denoting the heavier neutral Higgsino), the differential cross section is 
\begin{equation}\label{eq:diff_cross_section}
    \frac{d\sigma_{\chi A}}{dE_R}=\frac{\sigma_{\chi n} m_A}{2\mu_n^2w^2}\,[N-(1-4\sin^2\theta_W)Z]^2F^2_A(q^2) \, ,
\end{equation}
where $m_A$ is the nuclear mass, the momentum transfer $q = \sqrt{2m_AE_R}$ for nuclear recoil energy $E_R$, $\mu_n$ is the $\chi$-nucleon reduced mass $\simeq m_n$ for $m_\chi \gg m_n$, $F_A$ is the Helm form factor~\cite{Helm:1956zz}, and the $\chi$-nucleon cross-section $\sigma_{\chi n}=7.4\times10^{-39}~\rm cm^2$~\cite{Fan:2026kxx}. 
The allowed $E_R$ range is $E_R^{\pm}=(\mu^2_A/2m_A)(w\pm w')^2$ with $w'=\sqrt{w^2-2\delta/\mu_A}$ and $\mu_A$ is the $\chi$–nucleus reduced mass. 
For $\delta \gtrsim 80\rm\,keV$, the decay $\tilde{\chi} \rightarrow \chi + \gamma$ opens up with a decay length smaller than the solar radius~\cite{Fan:2026kxx}, hence the net effect is the capture of $\chi$. 

The summation in Eq.~\ref{eq:capture_rate} is over the solar elements.
We take the solar elemental composition from the present-day solar photospheric abundances in Table 1 of Ref.~\cite{Asplund:2004eu},
and the solar density profile from Ref.~\cite{Bahcall:2004pz}. 
We conservatively exclude elements such as uranium, for which only meteoritic ($\neq$ photospheric) abundances are tabulated in Ref.~\cite{Asplund:2021}. We take oxygen's radial density profile of Ref.~\cite{Bahcall:2004pz} as our benchmark and use
\begin{equation}
    n_A(r)=\rho_\odot(r)\left(\frac{X_O(r)}{A_O\,m_u}\right)10^{\log(\epsilon_A/\epsilon_O)}~,
    \label{eq:solarabundances}
\end{equation}
where $n_A(r)$ is the number density profile of species $A$,  $\rho_\odot(r)$ is the solar density profile, $X_O(r)$ and $A_O$ are the oxygen mass fraction and mass number,  $m_u$ is the atomic mass unit, and $\log\epsilon_A$ and $\log\epsilon_O$ are the photospheric logarithmic abundances of $A$ and oxygen.

\begin{figure*}[t]
    \centering
   \begin{subfigure}{0.495\textwidth}
        \centering
        \includegraphics[width=0.975\linewidth]{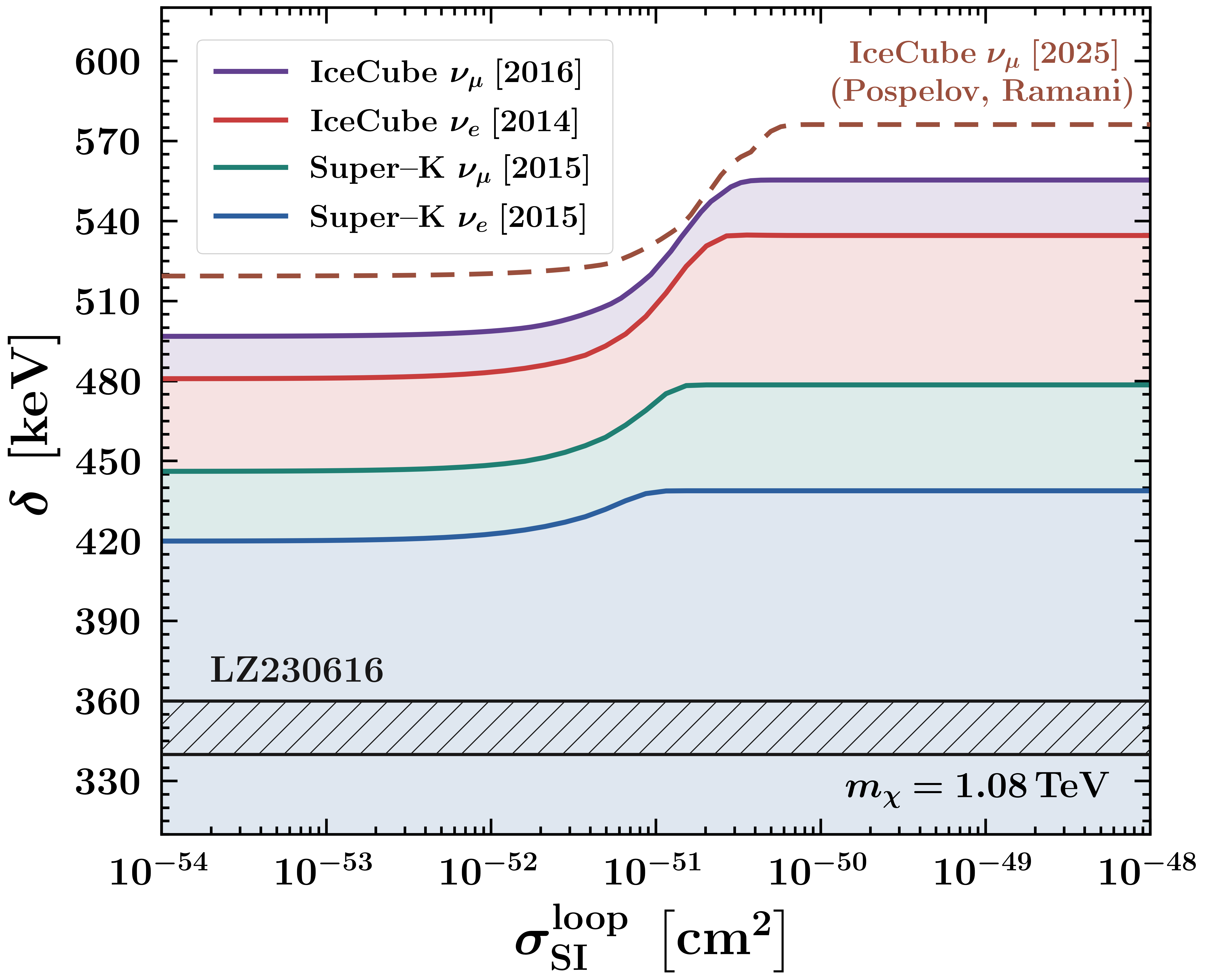}
     \end{subfigure}
    \begin{subfigure}{0.495\textwidth}
        \centering
        \includegraphics[width=0.975\linewidth]{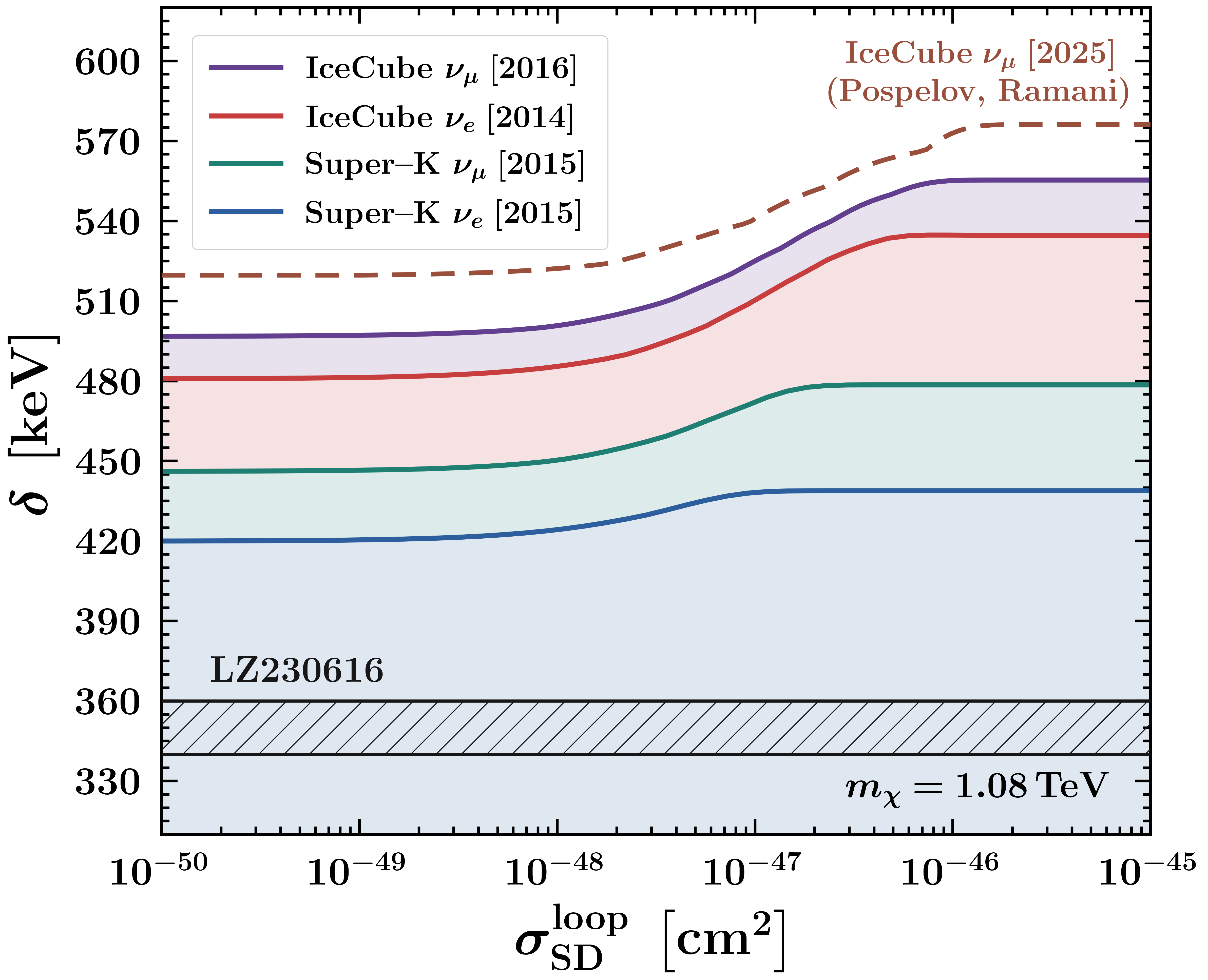}
    \end{subfigure}
%
    \caption{\justifying $95\%$ C.L. limits on the inter-state mass splitting as a function of loop-induced SI and SD elastic scattering cross sections for thermal Higgsino dark matter from solar $\nu_\mu$s at IceCube~\cite{IceCube:2016dgk}, atmospheric $\nu_e$s at IceCube~\cite{IceCube:2015mgt}, and atmospheric $\nu_e$s and $\nu_\mu$s at Super-K~\cite{Super-Kamiokande:2015qek}. 
     The grey band depicts the range of $\delta$ preferred by LZ230616~\cite{LZ:2026axp,Pospelov:2026ewn}. Also shown are IceCube-$\nu_\mu$ [2025] limits from Ref.~\cite{Pospelov:2026ewn}, with a 0.3$\to$0.4 rescaling of the DM density in GeV/cm$^3$.
     The vertical dashed line marks the mass of the thermal Higgsino, 1.08 TeV.
     See Sec.~\ref{sec:results} for more details.}
    \label{fig:higgsino_limits_SK_IceCube_nue_numu}
\end{figure*}

Following solar capture, the population of gravitationally bound DM particles evolves with time $t$ as
\begin{eqnarray}
\label{eq:dN_dt}
    \dfrac{d N_\chi}{d t} &=& C_\chi - \Gamma_{\rm ann} = C_\chi -  C_{\rm ann} N_\chi^2 \,\\
    \implies N_{\chi} (t) &=& \nonumber \sqrt{\dfrac{C_\chi}{C_{\rm ann}}} \, {\rm tanh} \left( \dfrac{t}{t_{\rm eq}} \right)~,\\
   \nonumber t_{\rm eq} &=& 1/\sqrt{C_\chi\, C_{\rm ann}}~,
\end{eqnarray}
where $C_{\rm ann}$ is the annihilation co-efficient with value derived below. For the $m_\chi$ considered here, evaporation is negligible~\cite{Garani:2021feo}. Following capture, $\chi$ repeatedly scatters on SM constituents, thermalizes within the Sun, and settles into a volume of radius $r_{\rm th}$. 
Repeated tree-level inelastic scatterings may become kinematically forbidden, and the captured $\chi$ may remain on extended orbits. 
However, loop-level elastic spin-independent (SI) and spin-dependent (SI) scatterings with cross sections $\sigma_{\rm SI}^{\rm loop}$ and $\sigma_{\rm SD}^{\rm loop}$ could then further shrink the orbits. 
The SI contribution receives an $A^2$ enhancement from the target nuclei, and due to accidental cancellations in the amplitude the cross section could be anywhere between zero and $3\times 10^{-49}$~cm$^2$~\cite{Chen:2019gtm}.
The SD scattering, for which we use only hydrogen targets, has cross section estimated to be about $5\times10^{-47}$~cm$^2$~\cite{Hisano:2011cs}.
For illustrating our results we float both $\sigma_{\rm SI}^{\rm loop}$ and $\sigma_{\rm SD}^{\rm loop}$ as free parameters.

Following Ref.~\cite{Pospelov:2026ewn}, we determine the orbital speed of captured DM from
\begin{equation}\label{eq:dr_dt}
    \dfrac{d r_{\chi, \rm orbit}}{dt} = \dfrac{r_{\chi, \rm orbit}^2}{G M(r_{\chi, \rm orbit})} \left\langle \dfrac{d \mathcal{E}_\chi}{dt} \right\rangle \, ,
\end{equation}
where $\mathcal{E}_\chi$ is the specific orbital energy of $\chi$. 
The averaged energy loss rate is estimated using the corresponding loop-level cross section and $n_A(r)$ from Eq.~\eqref{eq:solarabundances}. 
After evolving the orbit over the solar age $t_\odot = 4.5$~Gyr, the effective radius of the $\chi$ population is $ r_\chi(t_\odot) = {\rm max} \left( r_{\chi, \rm orbit} \, , \, r_{\rm th} \right)$. 
With this we obtain, for a thermally averaged cross section $\left\langle \sigma v \right\rangle_{\rm ann}$,
\begin{equation}\label{eq:Cann}
    C_{\rm ann} = \dfrac{\left\langle \sigma v \right\rangle_{\rm ann}}{\frac{4 \pi}{3} r_\chi^3(t_\odot)}~.
\end{equation}
For Higgsino DM, the captured $\chi$s chiefly annihilate to the $W^+W^-$ and $ZZ$ channels, producing high-energy secondary neutrinos. The differential neutrino flux at Earth is 
\begin{equation}\label{eq:dPhi_dE}
    \dfrac{d \Phi_{\nu + \bar{\nu}}}{dE_\nu} = \dfrac{\Gamma_{\rm ann}}{4 \pi ({\rm AU})^2} \sum_f {\rm Br}\left( \chi \chi \rightarrow f \right) \bigg[ \dfrac{dN_{\nu + \bar{\nu}}}{dE_\nu} \bigg]_{\chi \chi \rightarrow f} \, ,
\end{equation}
where $f$ denotes the annihilation channel with branching fraction ${\rm Br}(\chi \chi \rightarrow f)$ = 5/8 for $f=W^+W^-$ and 3/8 for $f=ZZ$ for thermal Higgsino DM~\cite{Dessert:2022evk, Rodd:2024qsi}, a benchmark we will also adopt for the non-thermal variant. 
Similarly, we use $\left\langle\sigma v\right\rangle_{\rm ann}=1.3\times10^{-26}~{\rm cm^3/s}$ -- appropriate for a $1.08~{\rm TeV}$ thermal Higgsino~\cite{Beneke:2014hja, Pospelov:2026ewn} -- but adopted for the non-thermal Higgsino as well. 
The neutrino spectrum $dN_{\nu + \bar{\nu}}/dE_\nu$ incorporates production and propagation effects. 
At production we compute it using $\texttt{$\chi$aro$\nu$}$~\cite{Liu:2020ckq}, which accounts for decays and interactions of the final state in the solar interior, along with electroweak corrections. 
We then propagate the resulting spectra from the solar core to terrestrial detectors using \texttt{nuSQuIDS}~\cite{Arguelles:2021twb}. 
We include neutrino interactions, flavor oscillations, and tau regeneration during propagation to obtain the spectra at Earth. 
In our region of interest, matter effects within Earth are negligible, the dominant matter effects arising within the solar interior.

\begin{figure*}[t]
    \centering
    \begin{subfigure}{0.495\textwidth}
        \centering
        \includegraphics[width=0.975\linewidth]{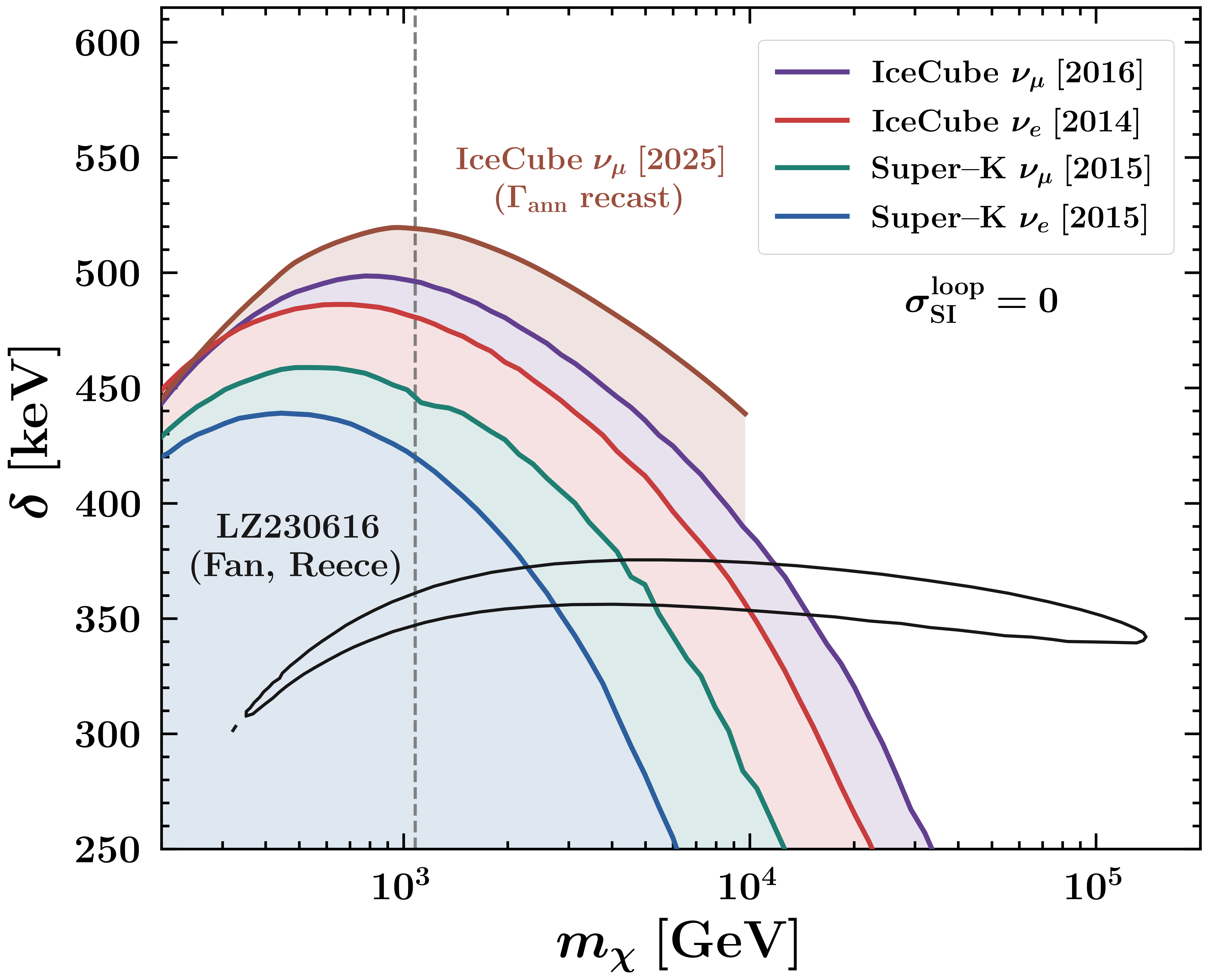}
        \label{sf:HLP_limits_SK_IceCube_nue_numu_sigma0}
    \end{subfigure}
    \begin{subfigure}{0.495\textwidth}
        \centering
        \includegraphics[width=0.975\linewidth]{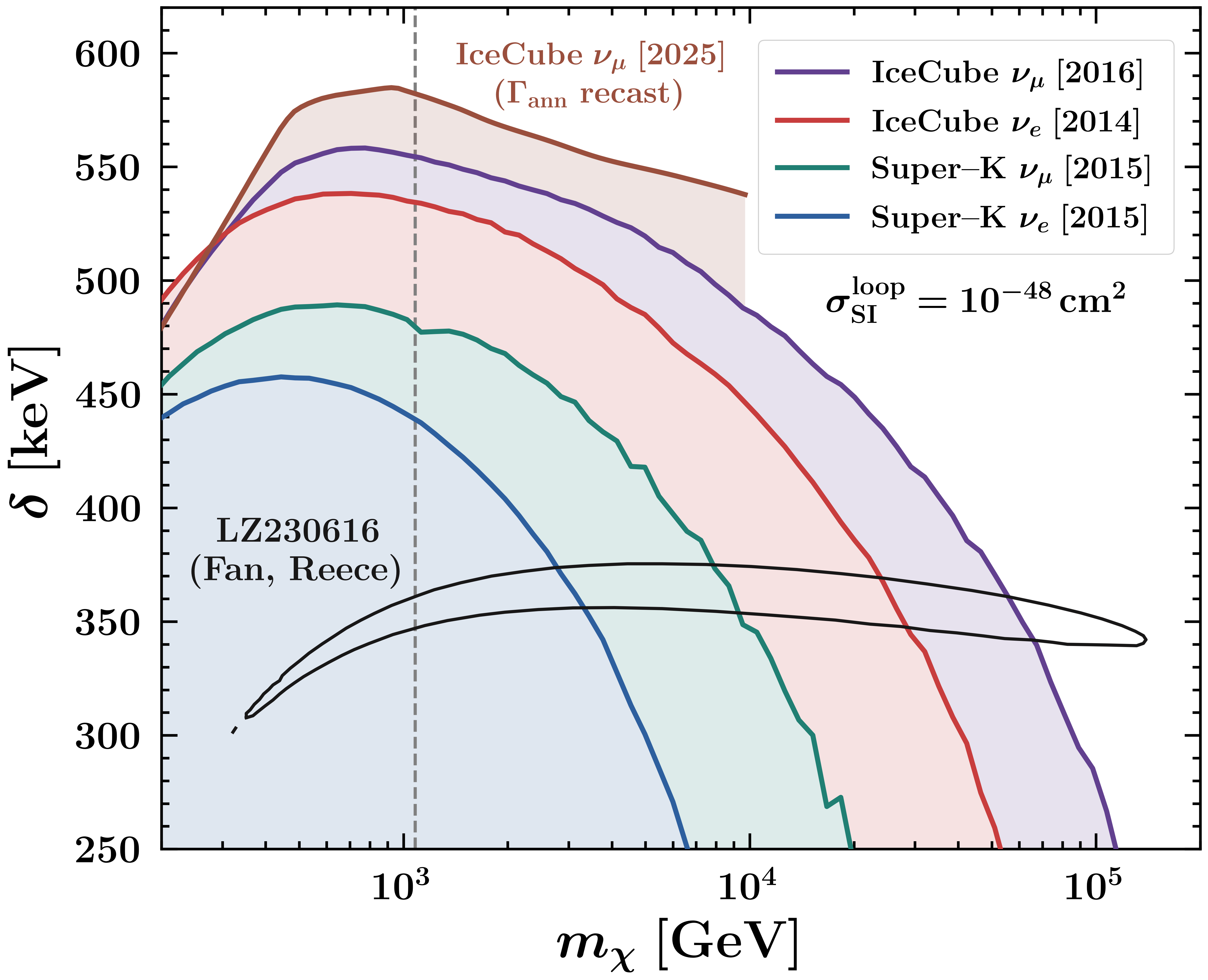}
        \label{sf:HLP_limits_SK_IceCube_nue_numu_SI_1e-48}
    \end{subfigure}
    \caption{\justifying
    95\% C.L. constraints on Higgsino-like DM, assuming $\sigma_{\rm SI}^{\rm loop} = 0$ (left panel) and $\sigma_{\rm SI}^{\rm loop} = 10^{-48} \, {\rm cm^2}$ (right panel), from the neutrino searches used in  Fig.~\ref{fig:higgsino_limits_SK_IceCube_nue_numu}. We also show the region that can explain LZ230616 assuming the Standard Halo Model~\cite{Fan:2026kxx}. Similar limits may be obtained for loop-induced spin-dependent scatters by choosing $\sigma_{\rm SD}^{\rm loop}$ at either end of its range in Fig.~\ref{fig:higgsino_limits_SK_IceCube_nue_numu}. See Sec.~\ref{sec:results} for more details.}
    \label{fig:HLP_limits_SK_IceCube_nue_numu}
\end{figure*}
%
%

\section{Analysis of datasets}
\label{sec:datasets}

Below we describe the IceCube and Super-K datasets we consider, using which we derive limits by computing
\begin{equation}\label{eq:chisq}
    \chi^2 = \sum_{i=1}^{N_{\rm bin}} \frac{\left(\Phi_{\rm d}^i - \Phi_{\rm atm}^i - \Phi_{\chi}^i\right)^2}{{\sigma_{\rm d}^i}^2} \, ,  
\end{equation}
where for the $i^{\rm th}$ bin, $\Phi_{\rm d}^i$ is the measured $\nu + {\bar \nu}$ flux with uncertainty $\sigma_{\rm d}^{i}$, $\Phi_{\rm atm}^i$ is the atmospheric neutrino model flux (with appropriate angular cuts), and $\Phi_{\chi}^i$ is the signal neutrino flux. We then obtain the 95\% C.L. upper limits on $(\delta,\, \sigma_{\rm SD}^{\rm loop})$ and $(\delta,\sigma_{\rm SI}^{\rm loop})$ by setting  $\chi^2-\chi_{\rm min}^2 = 2.71$, where $\chi_{\rm min}^2$ is the minimum  $\chi^2$ obtained by varying our model parameters~\cite{Lamperstorfer:2015cfg}.
For the IceCube\,--\,$\nu_\mu$ analysis, in Eq.\,\ref{eq:chisq} we use the predicted and observed {\em event numbers} rather than neutrino fluxes.

\textbf{IceCube\,--\,$\nu_\mu$ [2016]}~\cite{IceCube:2016dgk}, livetime $T=532$~days: 

High energy $\nu_\mu$s originating in the Sun can scatter in IceCube via charged currents and produce muon tracks. 
In this analysis IceCube focuses on up-going muon tracks from the solar direction to veto the much larger down-going atmospheric muon background, yet nevertheless picks up an irreducible background from Earth's atmospheric neutrinos.
Following Ref.\,\cite{Maity:2023rez}, we evaluate the differential event rate with respect to the solar opening angle $\theta_\odot$ as
\begin{equation}\label{eq:event_rate_theta}
\begin{aligned}
    \frac{dN_{\theta_\odot}}{d\cos \theta_\odot}
    &=
    2T
    \int_{E_\nu^{\min}}^{E_\nu^{\max}}
    A_{\rm eff}(E_\nu)
    \frac{d\Phi_{\nu +\bar{\nu}}}{dE_\nu}
    \frac{1}{\sqrt{2\pi}\sigma_\theta}
    \\
    &\qquad\times
    \exp\left[
    -\frac{\left(1-\cos \theta_\odot \right)^2}
    {2\sigma_\theta^2}
    \right]
    \, dE_\nu \,,
\end{aligned}
\end{equation}
where we adopt the energy-dependent effective area $A_{\rm eff} (E)$ from Ref.~\cite{IceCube:2016dgk} and the angular dispersion is
\begin{equation}
    \sigma_\theta
    =
    \left|
    \frac{
    \sqrt{2}\left(1-\cos\left[\Delta\theta(E_\nu)\right]\right)
    }{
    2\,{\rm erf^{-1}}(0.5)
    }
    \right| \,\,,
\end{equation}
where the median energy resolution $\Delta\theta(E_\nu)$ is taken from Ref.\,\cite{IceCube:2016dgk} and erf$^{-1}$ is the inverse error function. We compute the angular distribution of track-like events from our signal and compare it with the data and atmospheric background events presented in Fig.\,6 of Ref.\,\cite{IceCube:2016dgk}. In Fig.~\ref{fig:higgsino_IceCube_numu_equilibrium_Fig2.png} we show signal event counts in this channel for the thermal Higgsino parameters that fit LZ230616: from its exceeding the data (and background) event count across multiple bins it is clear that this interpretation is in tension with atmospheric neutrino flux data. This conclusion is true for all the datasets used in this work.

\textbf{IceCube\,--\,$\nu_e$ [2014]}~\cite{IceCube:2015mgt}, livetime $T$ = 332.3 days:

In this run, using a boosted decision tree for particle identification (PID-BDT), IceCube separated atmospheric $\nu_e$ events from the total cascade event samples. 
We use the atmospheric $\nu_e$ flux data in Fig.\,12 of Ref.\,\cite{IceCube:2015mgt}. 
For the background model we use the ``modified Honda'' flux that includes the extrapolation to higher energies of the original Honda
flux estimates\,\cite{PhysRevD.75.043006}. 
Cascade events have poorer angular resolutions than track events, thus to focus on the neutrinos from the Sun and eliminate atmospheric neutrino backgrounds from other directions, we apply an angular cut to the all-sky flux measurement provided in Ref.\,\cite{IceCube:2015mgt} by multiplying the per-steradian flux by 
\begin{eqnarray}\label{eq: angular cut}
    2\pi \int_{0}^{\delta\theta(E_\nu)}
    d\theta\,\sin\theta
    &=&
    4\pi \sin^2\left(
    \frac{\delta\theta(E_\nu)}{2}
    \right) \, ,
\end{eqnarray}
where $\delta\theta(E_\nu)$ is the energy-dependent angular resolution of cascade $\nu_e$ events given in Ref.\,\cite{IceCube:2015mgt}.

\textbf{Super-K\,--\,$\nu_\mu$ and $\nu_e$ [2015]}~\cite{Super-Kamiokande:2015qek}:

We use Super-Kamiokande data from Runs I\,--\,IV for atmospheric $\nu_\mu$ and $\nu_e$ flux measurements. Super-K classifies their events as fully contained (FC), partially contained (PC), and up-going muon events. In our analysis, we use the atmospheric $\nu_\mu$ and $\nu_e$ measurements presented in Fig.\,7 of Ref.~\cite{Super-Kamiokande:2015qek} along with the  HKKM11 atmospheric background model\,\cite{Honda:2006qj, PhysRevD.83.123001}. As in the IceCube $\nu_e$ cascade analysis, here we apply an angular cut defined in Eq.\,\ref{eq: angular cut} to the all-sky measurements from Ref.~\cite{Super-Kamiokande:2015qek}. Further details on the angular resolutions adopted for the Super-K atmospheric $\nu_\mu$ and $\nu_e$ observations can be found in Ref.~\cite{Krishna:2025ncv}.
%


\section{Results}
\label{sec:results}

In Fig.\,\ref{fig:higgsino_limits_SK_IceCube_nue_numu} we show for thermal Higgsino DM our limits, as well as a band enclosing $\delta = 340-360$~keV preferred by LZ230616.
The slope of these limits can be understood from the tree-level vs loop-level scattering of $\chi$. For the heaviest solar nuclei, tree-level inelastic scattering is suppressed due to kinematic threshold when $\chi$ reaches a radius $\sim 0.2 \,R_{\odot}$ after losing kinetic energy via repeated scatters. If the loop-level SI and SD cross sections are then large enough, $\chi$ can scatter further and settle into smaller volumes within the Sun, which would enhance its annihilation rates. This is reflected in Fig.\,\ref{fig:higgsino_limits_SK_IceCube_nue_numu}, where for SI (SD) cross sections up to $\sim 2\times 10^{-52}$ cm$^2$ ($\sim  2\times 10^{-48}$ cm$^2$) loop-level elastic scattering is inefficient for the thermalization of $\chi$, and the limit on $\delta$ is flat. As these cross sections are increased, $r_\chi$ shrinks and saturates to $r_{\rm th}$, where the limit on $\delta$ flattens again.

The hierarchy of our limits is understood from detector sizes, energy thresholds, and angular resolutions. The 1.08~TeV thermal Higgsino produces a broad neutrino spectrum in the Sun. With its higher energy threshold than Super-K, IceCube probes the harder part of the spectrum, containing a significant fraction of the signal flux. However, this flux is rapidly falling with energy, hence the smaller detector of Super-K registers much fewer events -- though thanks to its lower threshold it can access the softer part of the signal spectrum that IceCube cannot reach. 
Together, these effects make Super-K limits weaker than those from IceCube. Further, IceCube\,--\,$\nu_\mu$ measurements benefit from a significantly better angular resolution than the $\nu_e$ cascade observations, explaining its slightly stronger limits. Now between the Super-K datasets, although the atmospheric $\nu_\mu$ flux is smaller than that of $\nu_e$s, the $\nu_e$ data is only provided up to 100 GeV while $\nu_\mu$ measurements extend to higher energies, making the Super-K--$\nu_\mu$ limits stronger. Note that our strongest (IceCube\,--\,$\nu_\mu$) and weakest (Super-K\,--\,$\nu_e$) limits on $\delta$ differ by only $\sim 20\%$ despite various uncertainties in instrumental systematics and atmospheric neutrino models. 
Our limits overall are competitive with those obtained from the IceCube 2025 dataset by Ref.~\cite{Pospelov:2026ewn}.  

In Fig.\,\ref{fig:HLP_limits_SK_IceCube_nue_numu} we show our limits on the non-thermal Higgsino by varying $m_\chi$, also adding a recast of IceCube\,--\,$\nu_\mu$ [2025]\,\cite{IceCube:2025fcu} data for an indirect comparison with Ref.~\cite{Pospelov:2026ewn} (which only considered a thermal Higgsino). For the recast, we use the limits in Ref.\,\cite{IceCube:2025fcu} (provided up to $m_\chi$ = $10^4$ GeV) and scale it by the SI capture rate in Ref.\,\cite{Garani:2017jcj} to obtain the limits on the DM annihilation rate, $\Gamma_{\rm ann}$. The hierarchy of limits follow that of Fig.\,\ref{fig:higgsino_limits_SK_IceCube_nue_numu}; our benchmarks 
$\sigma^{\rm loop}_{\rm SI} =0$ ($\sigma^{\rm loop}_{\rm SI} =10^{-48} $~cm$^2$) are representative of all loop-level SI cross sections above (below) $\sigma_{\rm SI} \sim 10^{-50} $ cm$^2$ that saturate the upper limit on $\delta$ as seen in Fig.\,\ref{fig:higgsino_limits_SK_IceCube_nue_numu}. For sub-TeV masses we see that the limits from IceCube are weakening, which is due to its steep energy threshold, and indeed for $m_\chi \sim 100\,{\rm GeV}$ Super-K outdoes IceCube. 
For reference, we show a region enclosed by a black curve -- most of which is ruled out by our limits -- that explains LZ230616~\cite{Fan:2026kxx}. We remark that loop-level SD scattering leads to  limits qualitatively similar to SI scattering, though the strengthening and saturation of the limits occur for $\sigma^{\rm loop}_{\rm SI} \gtrsim  10^{-46} $~cm$^2$, as seen in Fig.\,\ref{fig:higgsino_limits_SK_IceCube_nue_numu}.
%


\section{Discussion}
\label{sec:discs}

Our results show that high-energy $\nu_e$ and $\nu_\mu$ measurements at IceCube and Super-K disfavor the Higgsino interpretation of the anomalous LZ230616 event. Our results complement Ref.\,\cite{Pospelov:2026ewn}, which used the latest IceCube measurements of $\nu_\mu$s from the solar direction. We have also ruled out a significant parametric region of a non-thermal Higgsino DM interpretation of LZ230616. 
Future observations by upcoming telescopes such as Hyper-Kamiokande and KM3NeT can further improve these bounds and potentially discover the existence of a (non-)thermal Higgsino -- as could direct detectors such as LZ, XENONnT, and PandaX-4T.

We re-state that our results do not completely rule out the thermal Higgsino interpretation of LZ230616. 
It could remain viable if the local DM velocity distribution receives non-standard contributions such as from the motion of the LMC~\cite{Fan:2026kxx} or unvirialized substructure, e.g., tidal streams~\cite{OHare:2014nxd, Maity:2022enp, Aggarwal:2024ngx}. 
Moreover, for $\sim$250~keV recoils of xenon the Helm form factor is in a region where it rapidly oscillates with $E_R$, so that its use may not be robust.

With numerous models being proposed involving inelastic scatters explaining LZ230616~\cite{Visinelli:2026kgt, Yin:2026jnn, Jeesun:2026vzo, Gu:2026vto, Smirnov:2026aqk, Unwin:2026rdp, Lou:2026idn, Su:2026rwz, Yamashita:2026ump, McCabe:2026crm, Chattopadhyay:2026ryw, deLima:2026shq} and hints of other high-recoil events fitted with inelastic and momentum-dependent elastic operators~\cite{An:2025bby}, these are exciting times for the dark matter hunter!
%

\section*{Acknowledgments}

We acknowledge Tarak Nath Maity for collaboration in the early stages of this work.
D.B. acknowledges the Council of Scientific and Industrial Research (CSIR), Government of India, for supporting his research under the Research Associateship program through grant no.\,\,09/0079(24106)/2025-EMR-I.  
D.J.D. acknowledges the financial support provided by the Ministry of Education (MoE), Government of India. 
R.S.\,\,acknowledges the University Grants Commission (UGC), Government of India, for financial support via the UGC-NET Senior Research Fellowship. 
A.S. acknowledges financial support from IISc. 
D.G. and J.D. acknowledges the financial support provided by the Ministry of Education (MoE), Government of India. 
S.B. acknowledges the Council of Scientific and Industrial Research (CSIR), Government of India, for supporting his research under the CSIR Junior/Senior Research Fellowship program through grant no. 09/0079(15488)/2022-EMR-I.
R.M. is supported by the ANRF National Postdoctoral Fellowship (NPDF) under Project
No. PDF/2025/001161.  
N.R. acknowledges support from the grant ANRF/ECRG/2024/000387/PMS and the Infosys Foundation, Bangalore.
R.L.\,\,acknowledges financial support from the institute start-up funds and ANRF for the grant no.\,\,ANRF/ARG/2025/005140/PS. 
%

\bibliography{refs}

\end{document}